# FOR OBJECTIVE CAUSAL INFERENCE, DESIGN TRUMPS ANALYSIS[1]

By Donald B. Rubin

*Harvard University*

For obtaining causal inferences that are objective, and therefore have the best chance of revealing scientific truths, carefully designed and executed randomized experiments are generally considered to be the gold standard. Observational studies, in contrast, are generally fraught with problems that compromise any claim for objectivity of the resulting causal inferences. The thesis here is that observational studies have to be carefully designed to approximate randomized experiments, in particular, without examining any final outcome data. Often a candidate data set will have to be rejected as inadequate because of lack of data on key covariates, or because of lack of overlap in the distributions of key covariates between treatment and control groups, often revealed by careful propensity score analyses. Sometimes the template for the approximating randomized experiment will have to be altered, and the use of principal stratification can be helpful in doing this. These issues are discussed and illustrated using the framework of potential outcomes to define causal effects, which greatly clarifies critical issues.

## 1. Randomized experiments versus observational studies.

1.1. *Historical dichotomy between randomized and nonrandomized studies for causal effects.* For may years, causal inference based on randomized experiments, as described, for example, in classic texts by Fisher (1935), Kempthorne (1952), Cochran and Cox (1950) and Cox (1958), was an entirely distinct endeavor than causal inference based on observational data sets, described, for example, in texts by Blalock (1964), Kenny (1979), Campbell and Stanley (1963), Cook and Campbell (1979), Rothman (1986),

Received May 2008; revised June 2008.
[1]This work was supported in part by NSF Grant SES-05-50887 and NIH Grant R01 DA023879-01.

Lilienfeld and Lilienfeld (1976), Maddala (1977) and Cochran (1983). This began to change in the 1970's when the use of potential outcomes, commonly used in the context of randomized experiments to define causal effects since Neyman (1923), was used to define causal effects in both randomized experiments and observational studies [Rubin (1974)]. This allowed the definition of assignment mechanisms [Rubin (1975)], with randomized experiments as special cases, thereby allowing both types of studies for causal effects to be considered within a common framework sometimes called the Rubin Causal Model [RCM–Holland (1986)]. In particular, the same underlying principles can be used to design both types of studies, and the thesis of this article is that for objective causal inference, those principles must be used.

1.2. *The appeal of randomized experiments for estimating causal effects.* For many years, most researchers have agreed that for drawing inferences about causal effects, classical randomized experiments, when feasible, are preferable to other methods [e.g., Cochran (1965)]. However, randomized experiments can be infeasible for a variety of ethical and other practical considerations, and the length of time we may have to wait for their answers can be too long to be helpful for impending decisions. Nevertheless, the possibility of conducting a randomized experiment should still be considered whenever a causal question arises, a point also made by Cochran (1965), which he attributed to earlier work by Dorn (1953).

Among the well-known reasons for this admiration for randomized experiments is the objectivity of the decisions for treatment assignment—the decision rules are explicit with a probability strictly between zero and one that each experimental unit will be exposed to either the treatment or control condition (for simplicity of exposition, this article will deal with the case of only two experimental conditions or exposures, called generically "treatment" and "control"). These unit-level probabilities, or propensity scores [Rosenbaum and Rubin (1983)], are known from the design of the experiment, and are all that are needed to obtain unbiased estimates of average treatment effects (i.e., the average effect of the treatment relative to control across all units). This unbiasedness property is suggestive of the powerful role that propensity scores play in causal effect estimation, even though unbiasedness is not an essential, or even always desirable, property of estimators.

Another reason why randomized experiments are so appealing, a reason, that is, of course not really distinct from their objectivity, is that they achieve, in expectation, "balance" on all pre-treatment-assignment variables (i.e., covariates), both measured and unmeasured. Balance here means that within well-defined subgroups of treatment and control units, the distributions of covariates differ only randomly between the treatment and control units.

A third feature of randomized experiments is that they are automatically designed without access to any outcome data of any kind; again, a feature not entirely distinct from the previous reasons. In this sense, randomized experiments are "prospective." When implemented according to a proper protocol, there is no way to obtain an answer that systematically favors treatment over control, or vice versa.

The theme of this article is that many of the appealing features of randomized experiments can and should be duplicated when designing observational comparative studies, that is, nonrandomized studies whose purpose is to obtain, as closely as possible, the same answer that would have been obtained in a randomized experiment comparing the same analogous treatment and control conditions in the same population. In this process of design, the usual models relating observed final outcome data to observed covariates and treatment indicators play no part, just as they do not in the design of a randomized experiment. The only models that are used relate treatment indicators to observed covariates.

1.3. *Observational studies as approximations of randomized experiments.* All statistical studies for causal effects are seeking the same type of answer, and real world randomized experiments and comparative observational studies do not form a dichotomy, but rather are on a continuum, from well-suited for drawing causal inferences to poorly suited. For example, a randomized experiment with medical patients in which 90% of them do not comply with their assignments and there are many unintended missing values due to patient dropout is quite possibly less likely to lead to correct inferences for causal inferences than a carefully conducted observational study with similar patients, with many covariates recorded that are relevant to well-understood reasons for the assignment of treatment versus control conditions, and with no unintended missing values.

The underlying theoretical perspective for the approach taken here was called the "Rubin Causal Model (RCM)" by Holland (1986) for a sequence of papers written in the 1970s [Rubin (1974, 1975, 1976a, 1977, 1978, 1979a, 1980)]. The RCM can be seen as having two essential parts, together called the "potential outcomes with assignment mechanism" perspective [Rubin (1990a), page 476], and a third optional part, which involves extensions to include Bayesian inference, only briefly mentioned here because our focus is on design, not analysis.

The first part of the RCM is conceptual, and it defines causal effects as comparisons of "potential outcomes" (defined in Section 2) under different treatment conditions on a common set of units. It is critical that this first part be carefully articulated if causal inferences are to provide meaningful guidance for practice. The second part concerns the explicit consideration of an "assignment mechanism." The assignment mechanism describes the process that led to some units being exposed to the treatment condition and other units being exposed to the control condition. The careful description and implementation of these two "design" steps is absolutely essential for drawing objective inferences for causal effects in practice, whether in randomized experiments or observational studies, yet the steps are often effectively ignored in observational studies relative to details of the methods of analysis for causal effects. One of the reasons for this misplaced emphasis may be that the importance of design in practice is often difficult to convey in the context of technical statistical articles, and, as is common in many academic fields, technical dexterity can be more valued than practical wisdom.

This article is an attempt to refocus workers in observational studies on the importance of design, where by "design" I mean all contemplating, collecting, organizing, and analyzing of data that takes place prior to seeing any outcome data. Thus, for example, design includes conceptualization of the study and analyses of covariate data used to create matched treated-control samples or to create subclasses, each with similar covariate distributions for the treated and control subsamples, as well as the specification of the primary analysis plan for the outcome data. However, any analysis that requires final outcome data to implement is not part of design. The same point has been emphasized in Rubin (2002, 2007) and the subsequent editorial by D'Agostino and D'Agostino (2007).

A brief review of the two essential parts of the RCM will be given in Section 2, which introduces terminology and notation; an encyclopedia entry review is given by Imbens and Rubin (2008a), a chapter length review is in Rubin (2008), and a full-length text from this perspective is Imbens and Rubin (2008b). Section 3 focuses on the assignment mechanism, the real or hypothetical rule used to assign treatments to the units, and on the importance of trying to reconstruct the hypothetical randomized experiment that led to the observed data, this reconstruction being conducted without examining any final outcome data in that observational data set.

Then Section 4 illustrates the design of an observational study using propensity scores and subclassification, first in the context of a classic single-covariate example from Cochran (1968) with one background covariate. Section 4 goes on to explain how propensity score methods allow the design of observational studies to be extended to cases with many covariates, first with an example comparing treatments for breast cancer to illustrate how this extension can be applied, and second, with a marketing example to illustrate the kind of balance on observed covariates that can be achieved in practice. Section 5 uses a Karolinska Institute example to illustrate a different point: that the same observational data set may be used to support two (or more) different templates for underlying randomized experiments, and one that may be far more plausible than the other. The concluding Section 6 briefly summarizes major points.

## 2. Brief review of the parts of the RCM relevant to design.

2.1. *Part one: units, treatments, potential outcomes.* Three basic concepts are used to define causal effects in the RCM. A unit is a physical object, for example, a patient, at a particular place and point in time, say, time $t$. A treatment is an action or intervention that can be initiated or withheld from that unit at $t$ (e.g., an anti-hypertensive drug, a job-training program); if the active treatment is withheld, we will say that the unit has been exposed to the control treatment. Associated with that unit are two potential outcomes at a future point in time, say, $t^* > t$: the value of some outcome measurements $Y$ (e.g., cholesterol level, income, possibly vector valued with more than one component) if the active treatment is given at $t, Y(1)$, and the value of $Y$ at the same future point in time if the control treatment is given at $t, Y(0)$. The causal effect of the treatment on that unit is defined to be the comparison of the treatment and control potential outcomes at $t^*$ (e.g., their difference, their ratio, the ratio of their squares). The times $t$ can vary from unit to unit in a population of $N$ units, but typically the intervals, $t^* - t$, are essentially constant across the $N$ units.

The full set of potential outcomes comprises all values of the outcome $Y$ that could be observed in some real or hypothetical experiment comparing the active treatment to the control treatment in a population of $N$ units. Under the "Stable Unit-Treatment Value Assumption (SUTVA)" [Rubin (1980, 1990a)], the full set of potential outcomes for two treatments and the population of $N$ units can be represented by an array with $N$ rows, one for each unit, and two "super" columns, one for $Y(0)$ and one for $Y(1)$, "super" in the sense that $Y$ can be multi-component. The fundamental problem facing causal inference [Holland (1986); Rubin (1978), Section 2.4] is that for the $i$th unit, only one of the potential outcomes for each unit, either $Y(0)$ or $Y(1)$, can ever be observed. In contrast to outcome variables, covariates are variables, $X$, that for each unit take the same value no matter which treatment is applied to the unit, such as quantities determined (e.g., measured) before treatments are assigned (e.g., age, pre-treatment blood pressure or pre-treatment education). The values of all these variables under SUTVA is the $N$ row array, $[X, Y(0), Y(1)]$, which is the object of causal inference called "the science."

A causal effect is, by definition, a comparison of treatment and control potential outcomes on a common set of units; for example, the average $Y(1)$ minus the average $Y(0)$ across all units, or the median log $Y(1)$ verses the median log $Y(0)$ for those units who are female between 31 and 35 years old, as indicated by their $X$ values, or the median $[\log Y(1) - \log Y(0)]$ for those units whose $Y(0)$ and $Y(1)$ values are both positive. It is critically important in practice to keep this definition firmly in mind.

This first part of the RCM is conceptual and can, and typically should, be conducted before seeing any data, especially before seeing any outcome data. It forces the conceptualization of causal questions in terms of real or hypothetical manipulations: "No causation without manipulation" [Rubin (1975)]. The formal use of potential outcomes to define unit-level causal effects is due to Neyman in 1923 [Rubin (1990a)] in the context of randomized experiments, and was a marvelously clarifying contribution. But evidently this notation was not formally extended to nonrandomized settings until Rubin (1974), as discussed in Rubin (1990a, 2005) and Imbens and Rubin (2008a, 2008b).

The intuitive idea behind the use of potential outcomes to define causal effects must be very old. Nevertheless, in the context of nonrandomized observational studies, prior to 1974 everyone appeared to use the "observed outcome" notation when discussing "formal" causal inference. More explicitly, letting $W$ be the column vector indicating the treatment assignments for the units ($W_i = 1$ if treated, $W_i = 0$ if control), the observed outcome notation replaces the array of potential outcomes $[Y(0), Y(1)]$ with $Y_{\mathrm{obs}}$, where the $i$th component of $Y_{\mathrm{obs}}$ is

$$Y_{\mathrm{obs},i} = W_i Y_i(1) + (1 - W_i) Y_i(0). \tag{2.1}$$

The observed outcome notation is inadequate in general, and can lead to serious errors—see, for example, the discussions in Holland and Rubin (1983) on Lord's paradox, and in Rubin (2005), where errors are explicated that Fisher made because (I believe) he eschewed the potential outcome notation. The essential problem with $Y_{\mathrm{obs}}$ is that it mixes up the science [i.e., $Y(0)$ and $Y(1)$] with what is done to learn about the science via the assignment of treatment conditions to the units (i.e., $W_i$).

2.2. *Part 2: the assignment mechanism.* The second part of the RCM is the formulation, or positing, of an assignment mechanism, which describes the reasons for the missing and observed values of $Y(0)$ and $Y(1)$ through a probability model for $W$ given the science:

$$\Pr(W|X, Y(0), Y(1)). \tag{2.2}$$

Although this general formulation, with the possible dependence of assignments on the yet to be observed potential outcomes, arose first in Rubin (1975), special cases were much discussed prior to that. For example, randomized experiments [Neyman (1923, 1990), Fisher (1925)] are "unconfounded" [Rubin (1990b)],

$$\Pr(W|X, Y(0), Y(1)) = \Pr(W|X), \tag{2.3}$$

and they are "probabilistic" in the sense that their unit level probabilities, or propensity scores $-e_i$, are bounded between 0 and 1:

$$0 < e_i < 1, \tag{2.4}$$

where

$$e_i \equiv \Pr(W_i = 1 | X_i). \tag{2.5}$$

When the assignment mechanism is both probabilistic [(2.4) and (2.5)] and unconfounded (2.3), then for all assignments $W$ that have positive probability, the assignment mechanism generally can be written as proportional to the product of the unit level propensity scores, which emphasizes the importance of propensity scores in design:

$$\Pr(W | X, Y(0), Y(1)) \propto \prod_{i=1}^{N} e_i \quad \text{or} \quad = 0. \tag{2.6}$$

The collection of propensity scores defined by (2.5) is the most basic ingredient of an unconfounded assignment mechanism because of (2.6), and its use for objectively designing observational studies will be developed and illustrated here, primarily in Section 4, but also in the context of a more complex design discussed in Section 5.

The term "propensity scores" was coined in Rosenbaum and Rubin (1983), where an assignment mechanism satisfying (2.4) and (2.5) is called "strongly ignorable," a stronger version of "ignorable" assignment mechanisms, coined in Rubin (1976a, 1978), which allows possible dependence on observed values of the potential outcomes, $Y_{\text{obs}}$ defined by (2.1), such as in a sequential experiment:

$$\Pr(W | X, Y(0), Y(1)) = \Pr(W | X, Y_{\text{obs}}).$$

But until Rubin (1975), randomized experiments were not defined using (2.3) and (2.4), which explicitly show such experiments' freedom from any dependence on observed or missing potential outcomes. Instead, randomized experiments were *described* in such a way that the assignments only depended on available covariates, and so implicitly did not involve the potential outcomes themselves. But explicit mathematical notation, like Neyman's, can be a major advance over implicit descriptions.

Other special versions of assignment mechanisms were also discussed prior to Rubin (1975, 1978), but without the benefit of explicit equations for the assignment mechanism showing possible dependence on the potential outcomes. For example, in economics, Roy (1951) described, without equations or notation, "self-optimizing" behavior where each unit chooses the treatment with the optimal outcome. And another well-known example from economics is Haavelmo's (1944) formulation of supply and demand behavior. But these and other formulations in economics and elsewhere did not use the notation of an assignment mechanism, nor did they have methods of statistical inference for causal effects based on the assignment mechanism. Instead, "regression" models were used to predict $Y_{\text{obs},i}$ from $X_i$ and $W_i$, with possible restrictions on some regression coefficients and/or on "error" terms. In these models certain regression coefficients (e.g., for $W_i$ or for interactions with $W_i$) were interpreted as causal effects; analogous approaches were used in other social sciences, as well as in epidemiology and medical research, and are still common. Such regression models were and are based on combined assumptions about the assignment mechanism and about the science, which were typically only vaguely explicated because they often were stated through restrictions on error terms, and therefore could, and sometimes did, lead to mistakes.

Inferential methods based only on the assumption of a randomized assignment mechanism were proposed by Fisher (1925) and described by Neyman (1923) and further developed by others [see Rubin (1990a) for some references]. The existence of these assignment-based methods, and their success in practice, documents that the model for the assignment mechanism is more fundamental for inference for causal effects than a model for the science. These methods lead to concepts such as unbiased estimation and asymptotic confidence intervals (due to Neyman), and $p$-values or significance levels for sharp null hypotheses (due to Fisher), all defined by the distribution of statistics (e.g., the difference of treatment and control sample means) induced by the assignment mechanism. In some contexts, such as the U.S. Food and Drug Administration's approval of a new drug, such assignment mechanism-based analyses are considered the gold standard for confirmatory inferences.

The third and final part of the RCM is optional; it involves specifying a full probability model for the science, the quantity being conditioned on in the assignment mechanism (2.2), and therefore treated as fixed in assignment-based approaches. This approach is Bayesian, and was developed by Rubin (1975, 1978) and further developed, for example, in Imbens and Rubin (1997) and in many other places. This can, in special simple cases, lead to the use of standard models, such as ordinary least squares regression models, but such models are generally not relevant to the design of observational studies.

Of course, there are other frameworks for causal inference besides mine, including ones where models have some relevance, but that is not the topic or focus of this article. The reader interested in various uses of models on the science $(X, Y(0), Y(1))$ can examine the text by Morgan and Winship (2007), which provides a fairly comprehensive discussion of different approaches. Also informative, but with an applied and prescriptive attitude, including some advice on design issues, is the text by Shadish, Cook and Campbell (2002).

## 3. Design observational studies to approximate randomized trials—general advice.

3.1. *Overview.* A crucial idea when trying to estimate causal effects from an observational dataset is to conceptualize the observational dataset as having arisen from a complex randomized experiment, where the rules used to assign the treatment conditions have been lost and must be reconstructed. There are various steps that I consider essential for designing an objective observational study. These will be described in this section and then illustrated in the remaining parts of this article. In practice, the steps are not always conducted in the order given below, but often they are, especially when facing a particular candidate data set.

3.2. *What was the hypothetical randomized experiment that led to the observed dataset?* As a consequence of our conceptualization of an observational study's data as having arisen from a hypothetical randomized experiment, the first activity is to think hard about that hypothetical experiment. To start, what exactly were the treatment conditions and what exactly were the outcome (or response) variables? Be aware that a particular observational dataset can often be conceptualized as having arisen from a variety of different hypothetical experiments with differing treatment and control conditions and possibly differing outcome variables. For example, a dataset with copious measurements of humans' prenatal exposures to exogenous agents, such as hormones or barbiturates [e.g., Rosenbaum and Rubin (1985), Reinisch et al. (1995)], could be proposed to have arisen from a randomized experiment on prenatal hormone exposure, or a randomized experiment on prenatal barbiturate exposure, or a randomized factorial experiment on both hormone and barbiturate exposure. But the investigator must be clear about the hypothetical experiment that is to be approximated by the observational data at hand. Running regression programs is no substitute for careful thinking, and providing tables summarizing computer output is no substitute for precise writing and careful interpretation.

3.3. *Are sample sizes in the dataset adequate?* If the step presented in Section 3.1 is successful in the limited sense that measurements of both treatment conditions and outcomes seem to be available or obtainable from descriptions of the observational dataset, the next step is to decide whether the sample sizes in this dataset are large enough to learn anything of interest. Here is where traditional power calculations are relevant; also extensions, for example, involving the ratios of sample sizes needed to obtain well-matched samples [Rubin (1976b), Section 5], are relevant, and should be considered before plunging ahead. Sometimes, the sample sizes will be small, but the data set is the only one available to address an important question. In such a case, it is legitimate to proceed, but efforts to create better data should be initiated.

If the available samples appear adequate, then the next step is to strip any final outcome measurements from the dataset. When designing a randomized experiment, we cannot look at any outcome measurements before doing the design, and this crucial feature of randomized experiments can be, and I believe must be, implemented when designing observational studies—outcome-free design is absolutely critical for objectivity. This point was made very strongly in Rubin (2007), but somewhat surprisingly, it was not emphasized much in older work, for example, in Cochran's work on observational studies as reviewed in Rubin (1984), or even in most of my subsequent work on matching summarized in Rubin (2006) prior to the mid-1990s. But I now firmly believe that it is critical to hide all outcome data until the design phase is complete. A subtlety here concerns "intermediate outcome data" discussed in Section 5, such as compliance measurements.

3.4. *Who are the decision makers for treatment assignment and what measurements were available to them*? The next step is to think very carefully about why some units (e.g., medical patients) received the active treatment condition (e.g., surgery) versus the control treatment condition (e.g., no surgery): Who were the decision makers and what rules did they use? In a randomized experiment, the randomized decision rules are explicitly written down (hopefully), and in any subsequent publication, the rules are likewise typically explicitly described. But with an observational study, we have to work much harder to describe and justify the hypothetical approximating randomized assignment mechanism. In common practice with observational data, however, this step is ignored, and replaced by descriptions of the regression programs used, which is entirely inadequate. What is needed is a description of critical information in the hypothetical randomized experiment and how it corresponds to the observed data.

For example, what were the background variables measured on the experimental units that were available to those making treatment decisions, whether observed in the current dataset or not? These variables will be called the "key covariates" for this study. Was there more than one decision maker, and if so, is it plausible that all decision makers used the same rule, or nearly so, to make their treatment decisions? If not, in what ways did the decision rules possibly vary? It is remarkable to me that so many published observational studies are totally silent on how the authors think that treatment conditions were assigned, yet this is the single most crucial feature that makes their observational studies inferior to randomized experiments.

3.5. *Are key covariates measured well*? Next, consider the existence and quality of the key covariates' measurements. If the key covariates are very poorly measured, or not even available in the dataset being examined, it is typically a wise choice to look elsewhere for data to use to study the causal question at hand. Sometimes surrogate variables can be found that are known to be highly correlated with unmeasured key covariates and can proxy for them. But no amount of fancy analysis can salvage an inadequate data base unless there is substantial scientific knowledge to support heroic assumptions. This is a lesson that many researchers seem to have difficulty learning. Often the dataset being used is so obviously deficient with respect to key covariates that it seems as if the researcher was committed to using that dataset no matter how deficient. And interactions and nonlinear terms should not be forgotten when considering covariates that may be key; for example, the assignment rules for medical treatments could differ for those with and without medical insurance.

3.6. *Can balance be achieved on key covariates*? The next step is to try to find subgroups (subclasses, or matched pairs) of treated and control units such that within a subgroup, the treated and control units appear to be balanced with respect to their distributions of key covariates. That is, within such a subgroup, the treated and control units should look as if they could have been randomly divided (usually not with equal probability) into treatment and control conditions. Often, it will not be possible to achieve such balance in an entirely satisfactory way. In that situation, we may have to restrict inferences to a subpopulation of units where such balance can be achieved, or we may even decide that with this dataset we cannot achieve balance with enough units to make the study worthwhile. If so, we should usually forgo using this dataset to address the causal question being considered. A related issue is that if there appear to be many decision makers using differing rules (e.g., different hospitals with different rules for when to give a more expensive drug rather than a generic version), then achieving this balance will be more difficult because different efforts to create balance will be required for the differing decision makers. This point will be clearer in the context of particular examples.

3.7. *The result.* These six steps combine to make for objective observational study design in the sense that the resultant designed study can be conceptualized as a hypothetical, approximating randomized block (or paired comparison) experiment, whose blocks (or matched pairs) are our balancing groups, and where the probabilities of treatment versus control assignment may vary relatively dramatically across the blocks. This statement does not mean the researcher who follows these steps will achieve an answer similar to the one that would have been found in the analogous randomized experiment, but at least the observational study has a chance of doing so, whereas if these steps are not followed, I believe that it is only blind luck that could lead to a similar answer as in the analogous randomized experiment.

Sometimes the design effort can be so extensive that a description of it, with no analyses of any outcome data, can be itself publishable. For a specific example on peer influence on smoking behaviors, see Langenskold and Rubin (2008).

## 4. Examples using propensity scores and subclassification.

4.1. *Classic example with one observed covariate.* The following very simple example is taken from Cochran (1968) classic article on subclassification in observational studies, which uses some smoking data to illustrate ideas. Let us suppose that we want to compare death rates (the outcome variable of primary interest) among smoking males in the U.S., where the treatment condition is considered cigarette smoking and the control condition is cigar and pipe smoking. There exists a very large dataset with the death rates of smoking males in the U.S., and it distinguishes between these two types of smokers. So far, so good, in that we have a dataset with $Y$ and treatment indicators, and it is very large. Now we strip this dataset of all outcome data; no survival (i.e., $Y$) data are left and are held out of sight until the design phase is complete.

Next we ask (in a simple minded way, because this is only an illustrative example), who is the decision maker for treatment versus control, and what are the key covariates used to make this decision? It is relatively obvious that the main decision maker is the individual male smoker. It is also relatively obvious that the dominant covariate used to make this decision is age—most smokers start in their teens, and most start by smoking cigarettes, not pipes or cigars. Some pipe and cigar smokers start in college, but many start later in life. Cigarette smokers tend to have a more uniform distribution of ages. Other possible candidate key covariates are education, socio-economic status, occupational status, income, and so forth, all of which tend to be correlated with age, so to illustrate, we focus on age as our only $X$ variable. Then our hypothetical randomized experiment starts with male smokers and randomly assigns them to cigarette or cigar/pipe smoking, where the propensity to be a cigarette smoker rather than a cigar/pipe smoker is viewed as a function of age. In this dataset, age is very well-measured. When we compare the age distribution of cigarette smokers and age distribution of cigar/pipe smokers in the U.S. in this dataset, we see that the former are younger, but that there is substantial overlap in the distributions. Before moving on to the next step, we should worry about how people in the hypothetical experiment who died prior to the assembling of the observational dataset are represented, but, for simplicity in this illustrative example, we will move on to the next step.

How do we create subgroups of treatment and control males with more similar distributions of age than is seen overall, in fact, so similar that we could believe that the data arose from a randomized block experiment? Cochran's example used subclassification. First, the smokers are divided at the overall median into young smokers and old smokers—two subclasses, and then divided into young, middle aged, and old smokers, each of these three subclasses being equal size, and so forth. Finally, nine subclasses are used. The age distributions within each of the nine subclasses are very similar for the treatment condition and the control condition, just as if the men had been randomly assigned within the age subclasses to treatment and control, because there is such a narrow range of ages within each of the nine subclasses. And of great importance, there do exist both treatment and control males in each of nine subclasses.

The design phase can be considered complete for our simple illustrative example. Our underlying hypothetical randomized experiment that led to the observed dataset is a randomized block experiment with nine blocks defined by age, where the probability of being assigned to the treatment condition (cigarette smoking) rather than the control condition (cigar/pipe smoking) decreases with age. We are now allowed to look at the outcome data within each subclass and compare treatment and control death rates. We find that, averaging over the nine blocks (subclasses), the death rates are about 50% greater for the cigarette smokers than the cigar and pipe smokers. Incidentally, the full data set with no subclassification leads to nearly the opposite conclusion; see Cochran (1968) or Rubin (1997) for details.

But what would have happened if we decided that we wanted to subclassify also on education, socio-economic status, and income, each covariate using, let's say, five levels [a minimum number implicitly recommended in Cochran (1968)]? There would be four key covariates, each with five levels, yielding a total of 625 subclasses. And many observational studies have many more than four key covariates that are known to be used for making treatment decisions. For example, with 20 such covariates, even if each is dichotomous, there are $2^{20}$ subclasses—greater than a million, and as a result, many subclasses would probably have only one unit, either a treated or control, with no treatment-control comparison possible. How should we design this step of observational studies in such more realistic situations?

4.2. *Propensity score methodology.* Rosenbaum and Rubin (1983) proposed a class of methods to try to achieve balance in observational studies when there are many key covariates present. In recent years there has been an explosion of work on and interest in these methods; the Introduction in Rubin (2006) offers some references. Sadly, many of the articles that use propensity score methods do not use them correctly to help design observational studies according to the guidelines in Section 3, which are motivated by the theoretical perspective of Section 2 and illustrated in the trivial one-covariate example of Section 4.1. Rather, these inappropriate applications, for example, use the outcome data to help choose propensity score models, and use the propensity score only as a predictor in a regression model with the outcome, $Y_{\text{obs}}$, as the dependent variable.

The propensity score is the observational study analogue of complete randomization in a randomized experiment in the sense that its use is not intended to increase precision but only to eliminate systematic biases in treatment-control comparisons. In some cases, however, its use can increase precision; for the reason, see Rubin and Thomas (1992). As we have seen in earlier sections, it is formally defined as the probability of a unit receiving the treatment condition, rather than the control condition, as a function of observed covariates, including indicator variables for the individual decision makers and associated interactions, if needed. The propensity score is rarely known in an observational study, and therefore must be estimated, typically using a model such as logistic regression, but this choice, although common, is by no means mandatory or even ideal in many circumstances. The critical aspect of the propensity score is that it models the reasons for treatment versus control assignment at the level of the decision maker. For instance, in the context of the expanded tobacco example of Section 4.1, it could model the choice of a male smoker to smoke cigarettes versus cigars or pipes as a function of age, income, education, SES, etc. Once estimated, the linear version of it (e.g., the beta times $X$ in the logistic regression) can be treated as the only covariate, just like age in the example of Section 4.1, and it is used to match or subclassify the treatment and control units.

But we are not done yet. We have to check that balance on all covariates has been achieved. If the propensity score is correctly estimated and there is balance on it, then Rosenbaum and Rubin (1983) showed that balance is achieved on all observed covariates. The achieved balance within matched pairs or subclasses must be assessed and documented before the design phase is finished. With only one covariate, balance on that covariate is easily achieved (if it can be achieved) by using narrow enough subclasses (or bins) of the covariate. With many covariates, the assessment and re-estimation of propensity score to achieve balance can be tricky, and good guidance for doing this is still being developed. When selecting matched pairs, using both the propensity score and some prognostically important functions of key covariates can often result in increased precision of estimation [see Rubin (1979b), Rosenbaum and Rubin (1985), Rubin and Thomas (2000)].

Here we illustrate these various ideas in the context of some real examples. The next example concerns the relative success of two treatments for breast cancer, and illustrates not only the process of selecting the key background variables for use in the propensity score estimation, but also illustrates that careful observational studies can (not necessarily will) reach the same general conclusions as expensive randomized experiments. The second example is from a large marketing study and displays the kind of balance that can be achieved following propensity score subclassification, as well as the fact that some units can be unmatchable. The last example, in Section 5, uses a data set on large volume versus small volume hospitals to emphasize that one observational data set can be used to support two (or more) differing templates for the underlying randomized study of a particular question, and one template may be considered far better than the other.

4.3. *GAO study of treatments for breast cancer.* The following example appeared in a Government Accounting Office (GAO) publication that was summarized in Rubin (1997). In the 1960s mastectomy was the standard treatment for many forms of breast cancer, but there was growing interest in the possibility that for a class of less severe situations (e.g., small tumors, node negative) a more limited surgery, which just removed the tumor, might be just as successful as the more radical and disfiguring operation.

Several large and expensive randomized trials were done for this category of women with less severe cancer, and the results of these trials are summarized in Table 1. As can be seen there, these studies suggest that for this class of women who are willing to participate in a randomized experiment, and for these cancer treating centers and their doctors, who are also willing

TABLE 1

*Estimated 5-year survival rates for nodenegative patients in six randomized clinical trials*

| **Study** | **Women** | **Women** | **Estimated survival rate for women** | **Estimated survival rate for women** | **Estimated causal effect** |
|---|---|---|---|---|---|
| **Study** | **Breast conservation (BC)** | **Mastectomy (Mas)** | **BC** | **Mas** | **BC–Mas** |
| **Study** | ***n*** | ***n*** | **%** | **%** | **%** |
| U.S.–NCI† | 74 | 67 | 93.9 | 94.7 | −0.8 |
| Milanese† | 257 | 263 | 93.5 | 93.0 | 0.5 |
| French† | 59 | 62 | 94.9 | 96.2 | −1.3 |
| Danish‡ | 289 | 288 | 87.4 | 85.9 | 1.5 |
| EORTC‡ | 238 | 237 | 89.0 | 90.0 | −1.0 |
| U.S.–NSABP‡ | 330 | 309 | 89.0 | 88.0 | 1.0 |

† Single-center trial; ‡ Multicenter trial.

Reference: Rubin, D. B. Estimated causal effects from large datasets using propensity scores. Annals of Internal Medicine (1997); 127, 8(II):757–763.

to participate, the five-year survival rate appears to be very similar in the two randomized treatment conditions. There is, however, an indication from Table 1 that the survival is better overall in the trials conducted in single centers (the top three rows) than in the multi-center trials (the bottom three rows), possibly because of more specialized care, including after care.

The reason this last comment is relevant is that based on these results, the U.S. National Cancer Institute felt that for this category of women, the recommendation for general practice should be to have breast conserving operations rather than the radical version. The GAO was concerned that this advice based on these randomized trials may not be wise for general practice, where the surgeons involved may not be as skilled, after care may be lower quality, the women themselves may be less research-oriented and therefore less medically astute about their own care, and so forth, than in the randomized trials. It was not possible to initiate a new randomized trial in the general population of women and doctors who may not want to be randomized; even if it were, the funding, planning, implementing, etc., would take too long and results concerning five-year survival would be first available a decade in the future.

Consequently, the GAO implemented an observational study using the SEER (Surveillance, Epidemiology, End Results) data base, which has relatively complete information on all cancer cases in certain catchment areas. Importantly, it had detailed information on the kind and diagnosed severity of the cancer, so that they could use the same exclusion criteria as the randomized experiments, and it had the kind of treatment used by the surgical team; also, it had survival information, which of course was the key outcome variable. Moreover, it had many covariates. And it had about five thousand breast cancer cases of the type studied in the six randomized experiments during the relevant period, which was considered a large enough sample to proceed.

So far so good. The outcome data were stripped from the files, and the design phase proceeded. The following description is from an over 15 year old memory, and no doubt is somewhat distorted by my current attitudes, but is largely accurate, I believe. The GAO checked with a variety of physicians about who the decision makers were for choice of surgery for this category of women. The replies were that they were usually joint choices made by the surgeon and woman, sometimes with input from the husband or other family members or friends. Some of the key covariates were obvious, such as the size of the tumor and the woman's age and marital status. Others were less obvious, such as urbanization, region of the country, year, race, and various interactions (e.g., age by marital status). In any case, a list of approximately twenty key covariates was assembled, and it turned out that all had been collected in SEER. More good news. Then the consistency of the decision makers' rules across the dataset was considered, although at the time, not as seriously as I would do it now. It was decided that the way women and doctors used the key covariates was pretty much the same around the country, and any differences were probably captured by the observed covariates.

Propensity scores were estimated by logistic regression, and they were used to create five subclasses of treatment/control women. The women were ranked by their estimated propensity scores, and the lowest 20% formed subclass 1, the next 20% formed subclass 2, etc. Within each subclass, balance was checked, not only on the covariates included in the propensity score, but also on all other important covariates in the database. For example, the average age of a treated women within each subclass should be approximately the same as the average age of a control women in that subclass, and the proportion of each that are married should also be as similar as if the treatment and control women in that subclass had been randomly divided (obviously, not with equal probability across the subclasses). When less balance was found on a key covariate within a subclass than would have occurred in a randomized experiment, terms were added to the propensity score model and balance was reassessed. Unfortunately, those tables and the processes never survived into the final report, but such balance was achieved—not perfectly, but close enough to believe in the hypothetical underlying randomized block experiment that led to the observed data.

The results of the subclassification on the propensity score are summarized in Table 2. In general, this observational study's results are consistent with those from the randomized trials. There is essentially no evidence for any advantage to the radical operation, except possibly in those propensity score subclasses where the women and doctors were more likely to select

TABLE 2

*Estimated 5-year survival rates for node-negative patients in SEER data base within each of five propensity score subclasses: from tables in U.S. GAO Report [General Accounting Office (1994)]*

| **Propensity score subclass** | **Treatment condition** | ***n*** | **Estimate** |
|---|---|---|---|
| 1 | Brest conservation | 56 | 85.6% |
| | Mastectomy | 1008 | 86.7% |
| 2 | Brest conservation | 106 | 82.8% |
| | Mastectomy | 964 | 83.4% |
| 3 | Brest conservation | 193 | 85.2% |
| | Mastectomy | 866 | 88.8% |
| 4 | Brest conservation | 289 | 88.7% |
| | Mastectomy | 978 | 87.3% |
| 5 | Brest conservation | 462 | 89.0% |
| | Mastectomy | 604 | 88.5% |

mastectomy (subclasses 1, 2, 3), but the data are certainly not definitive. Similarly, for the women and doctors relatively more likely to select breast conserving operations, there is some slight evidence of a survival benefit to that choice. If we believed that the treatment effect should be the same for all women in the study, these changing results across propensity subclasses could be viewed as evidence of a confounded and nonignorable treatment assignment (i.e., an omitted key covariate). Overall, however, there appears to be no advantage to recommending one treatment over the other. It is interesting to note that, consistent with expectations, the overall survival rates in the observational dataset are not as good as those in the more specialized centers represented in Table 1.

4.4. *Marketing example.* Propensity score methods, like randomization, work best in large samples. For a trivial example, if we have one man and one woman, one to be treated and one to be control, randomized treatment assignment in expectation would create a half-man treated and half-woman control, but in reality the man would be either treated or control and the woman would be in the other condition. With a hundred men and a hundred women, we would expect roughly half of each to be in each treatment arm. Analogously, with propensity scores, the creation of narrow subclasses or matched pairs should create balanced distributions in expectation, which should be easier to achieve and to assess in large samples than in small ones.

The next example, from Rubin and Waterman (2006), illustrates this feature well because the sample sizes are large: 100,000 treated doctors and 150,000 control doctors. "Treated" here means visited by a sales representative (rep) at least once during a certain six month period; the sales rep tells the doctor the details of a new weight-loss drug being promoted by a pharmaceutical company. The control doctors are not visited by a sales rep from that company during that period. The treatment/control indicator variables come from the companies' records as provided by the sales reps. The key outcome variable is the number of prescriptions (scripts) of this drug written by the doctor during the following six months; this information on scripts is obtained several months later from a third party vendor, which is updated at regular intervals. The previous version of this data source and other sources have all sorts of background information on the doctors, such as sex, race, age, years since degree, size of practice, medical specialty, number of scripts written in prior years for the same class of drugs as being described by the sales rep, etc. In fact, there are well over 100 basic covariates available. The objective of the observational study is to estimate the causal effects of the reps visiting these doctors. It costs money to pay the sales reps' salaries to visit the doctors, and moreover, many reps get commissions based, in part, on the number of scripts written by the doctors they visited for the detailed drug. Do the visits cause more scripts to be written, and if so, which doctors should be visited with higher priority? Both of these, and other similar questions, are causal ones.

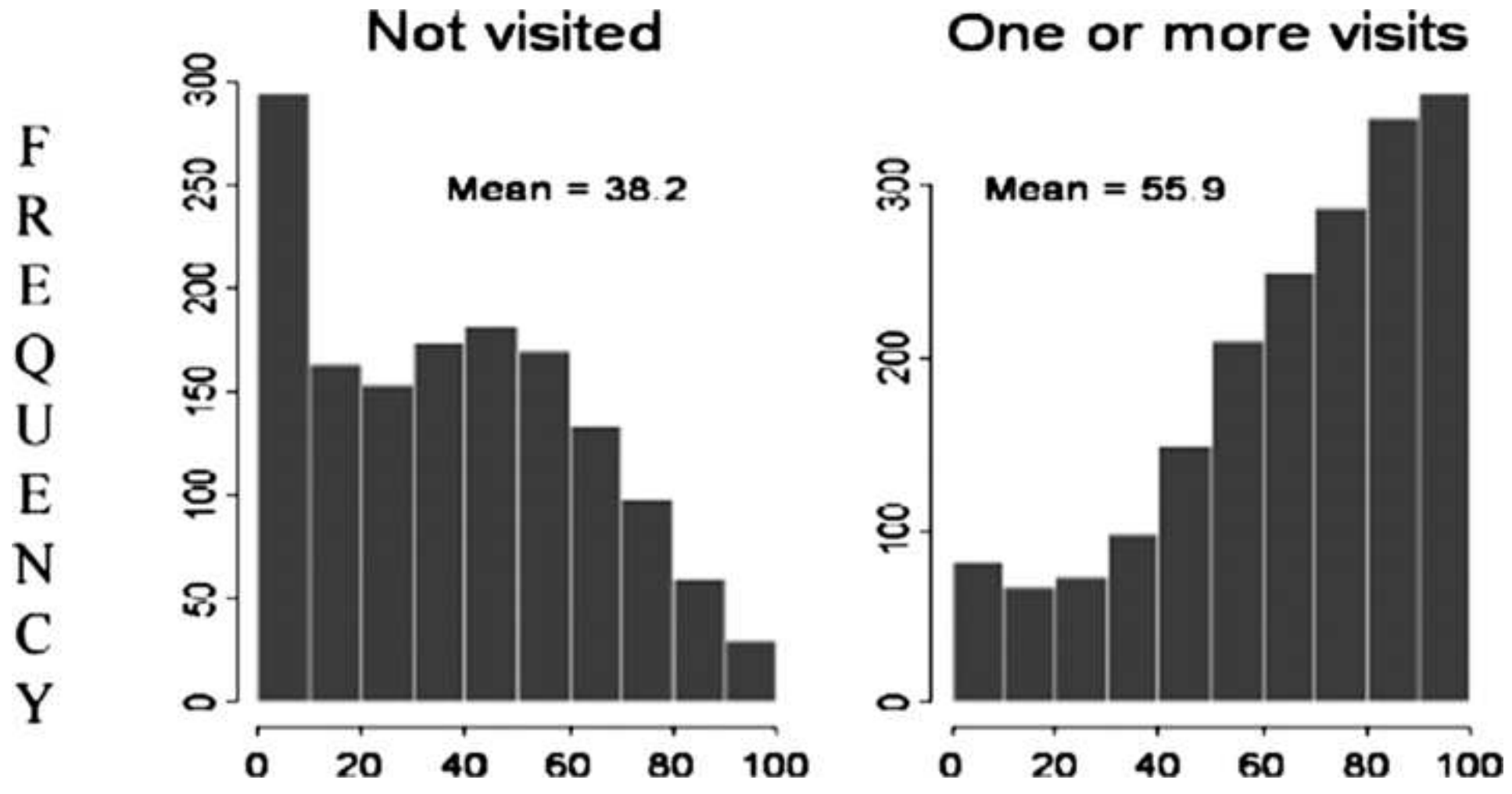

FIG. 1. *Histograms for background variable: prior Rx score (0–100) at baseline.*

The decision-maker for visiting or not the doctors is essentially the sales rep, and these folks, rather obviously, like to visit doctors who prescribe a lot, who have large practices, are in a specialty that prescribes a lot of the type of drug being detailed, etc. Essentially all of these background variables, $X$, and more, are available on the purchased data set, which has huge sample sizes; the company has the indicator $W$ for visited versus not, and next year's purchased data set will have the outcome variables $Y$ on the actual number of scripts written by these doctors in the next time period. So things look in good shape to estimate and re-estimate the propensity scores until we achieve balanced distributions within subclasses, or we decide that there are some types of doctors who have essentially no chance of being visited or not being visited, and then no estimation of causal effects will be attempted for them.

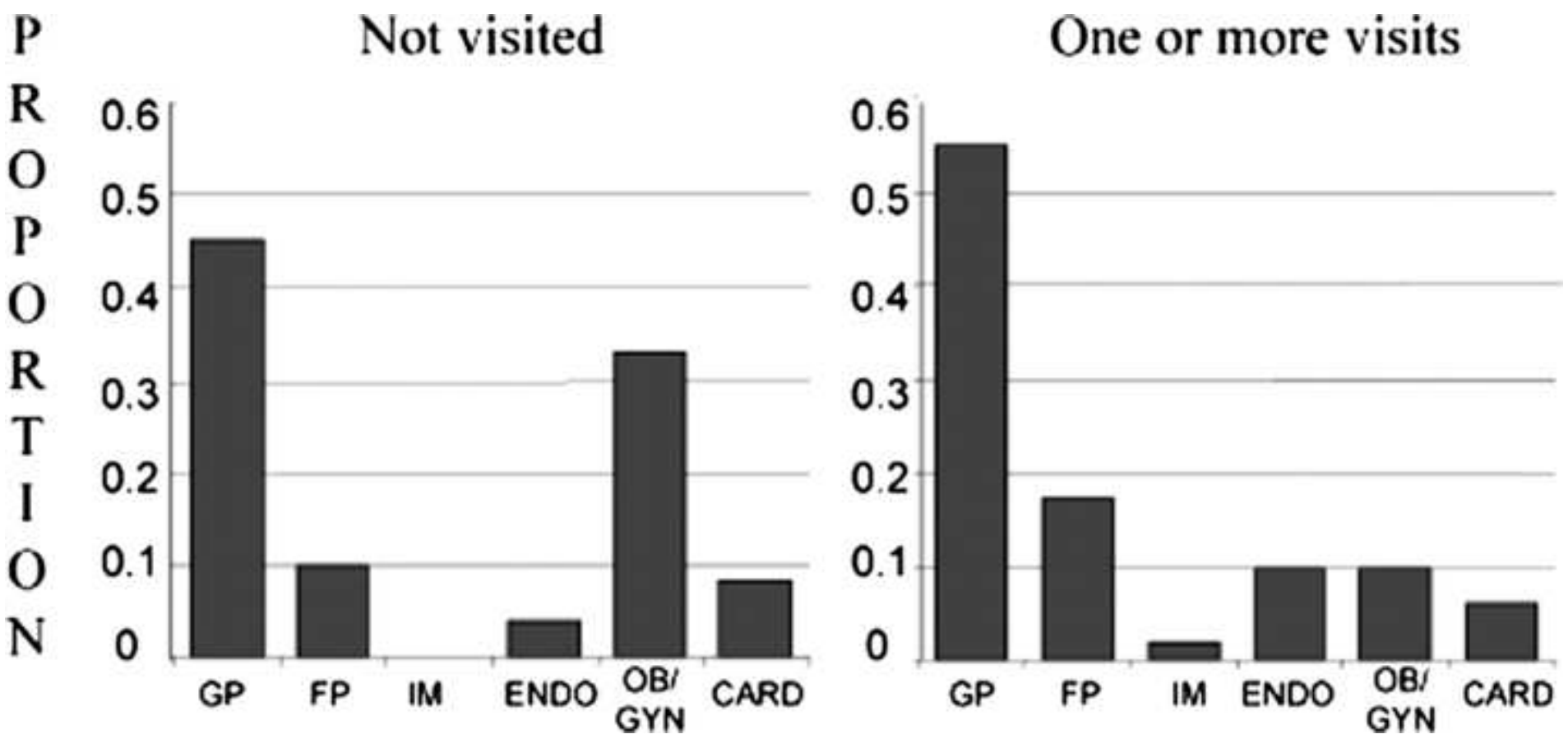

FIG. 2. *Histograms for background variable: specialty.*

Figures 1 and 2 display the initial balance for two important covariates, number of prior scripts written in the previous year (for drugs in the same class as the detailed drug) on a scale from 0 (minimum) to 100 (the arbitrarily scaled maximum), and the specialty. These figures reveal quite dramatic differences between the doctors who were visited and those who were not visited. It is not surprising that the visited doctors were the ones who wrote many more prescriptions (per doctor) than the not visited doctors. But the visited doctors also have a different distribution of specialties than the not visited doctors. For example, ob-gyn doctors are visited relatively less often than doctors with other specialties; presumably, ob-gyn doctors do not prescribe weight-loss drugs for their pregnant patients, and the sales reps use this information.

Propensity scores were estimated by logistic regression based on various functions of all of the covariates. Figure 3 displays the histograms for the estimated linear propensity scores (the $\hat{\beta}X$ in the logistic regression) among the not visited and visited doctors. These histograms are shown with 15 subclasses (or bins) of propensity scores. In some bins, there are only visited doctors, that is, in the bins with linear propensity scores larger than 1.0; in those two bins, there are no doctors who were not visited. Presumably, they are high prescribing doctors with large practices, etc. No causal inferences are possible for them without making model-based assumptions relating outcomes to covariates for which there are no data to assess the underlying assumptions. Similarly, for the four lowest bins of propensity, with linear scores less than 0.1, all doctors are not visited, and so, similarly, no causal inferences about the effect of visiting this type of doctor are possible unless based on unassessable assumptions.

But in the other nine bins, there are both visited and not visited doctors, and the claim is that within each of those bins, the distributions of all covariates that entered the propensity score estimation will be nearly the same for the visited and not visited doctors. To be specific, let us examine the bin between 0.5 and $<0.6$. Figures 4 and 5 show the distributions of prior number of prescriptions and specialties in this bin for the not visited and visited doctors. These distributions are strikingly more similar than their counterparts shown in Figures 1 and 2. In fact, they are so similar that one could believe that, within that bin, the visited doctors are a random sample from all doctors in that bin. And the claim is that this will hold (in expectation) for all covariates used to estimate the propensity score and in all bins where there are both visited and not visited doctors.

The process of assessing balance was conducted for all variables and all bins and considered adequate in the sense that it was considered plausible that a randomized block experiment had become reconstructed, except for the bins with only visited or not visited doctors. Admittedly, there is an aspect of “art” operating here, in that random imbalance of prognostically important covariates (i.e., ones thought to be strongly related to outcome variables) was considered more important to correct than more extreme imbalance in prognostically unimportant ones, but the field of statistics will always benefit from scientifically informed thought. Nevertheless, better guidance on how to conduct this process more systematically is needed, and is in development; see, for example, Imbens and Rubin [(2008b), Chapters 13 and 14].

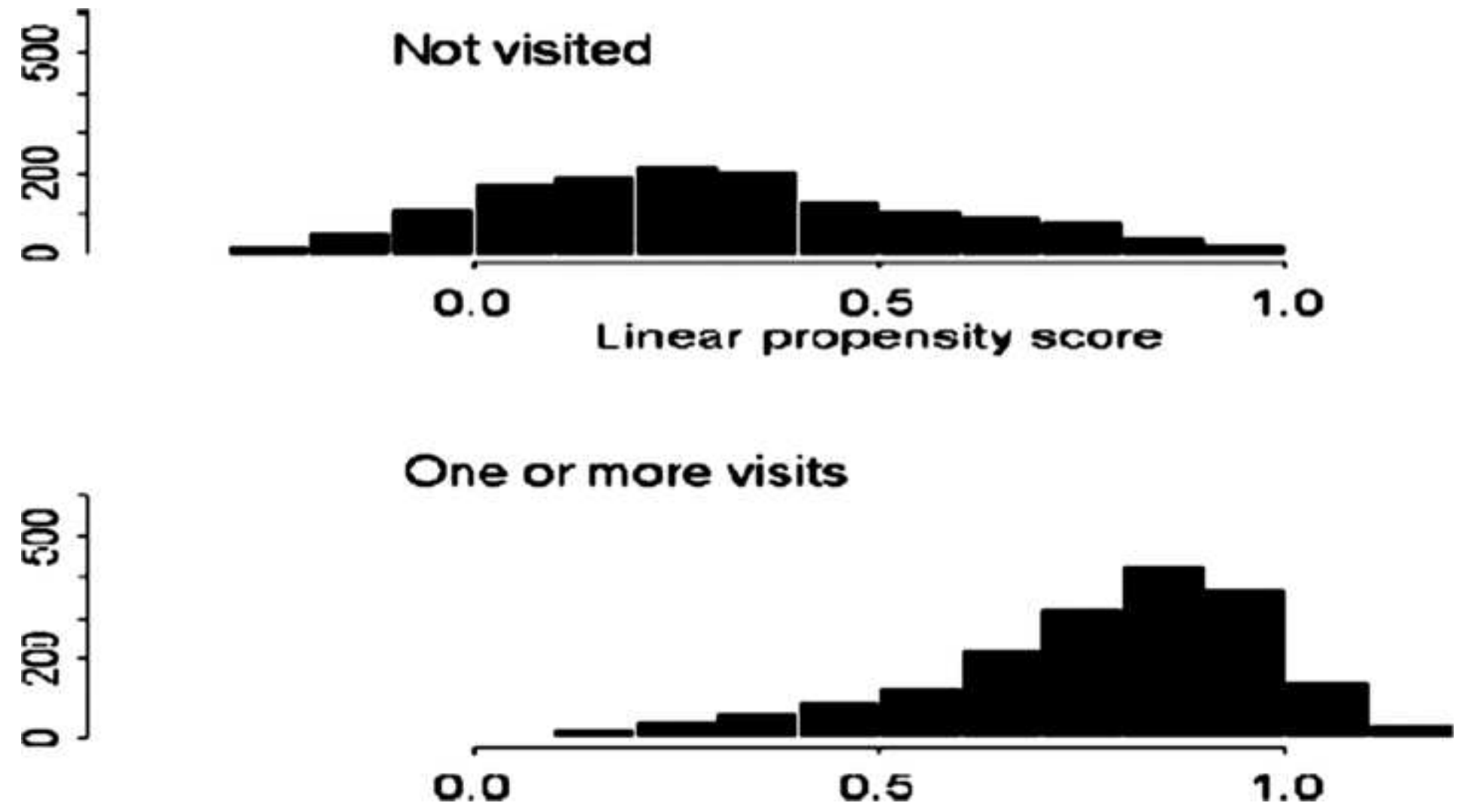

Fig. 3. *Histograms for summarized background variables: linear propensity score.*

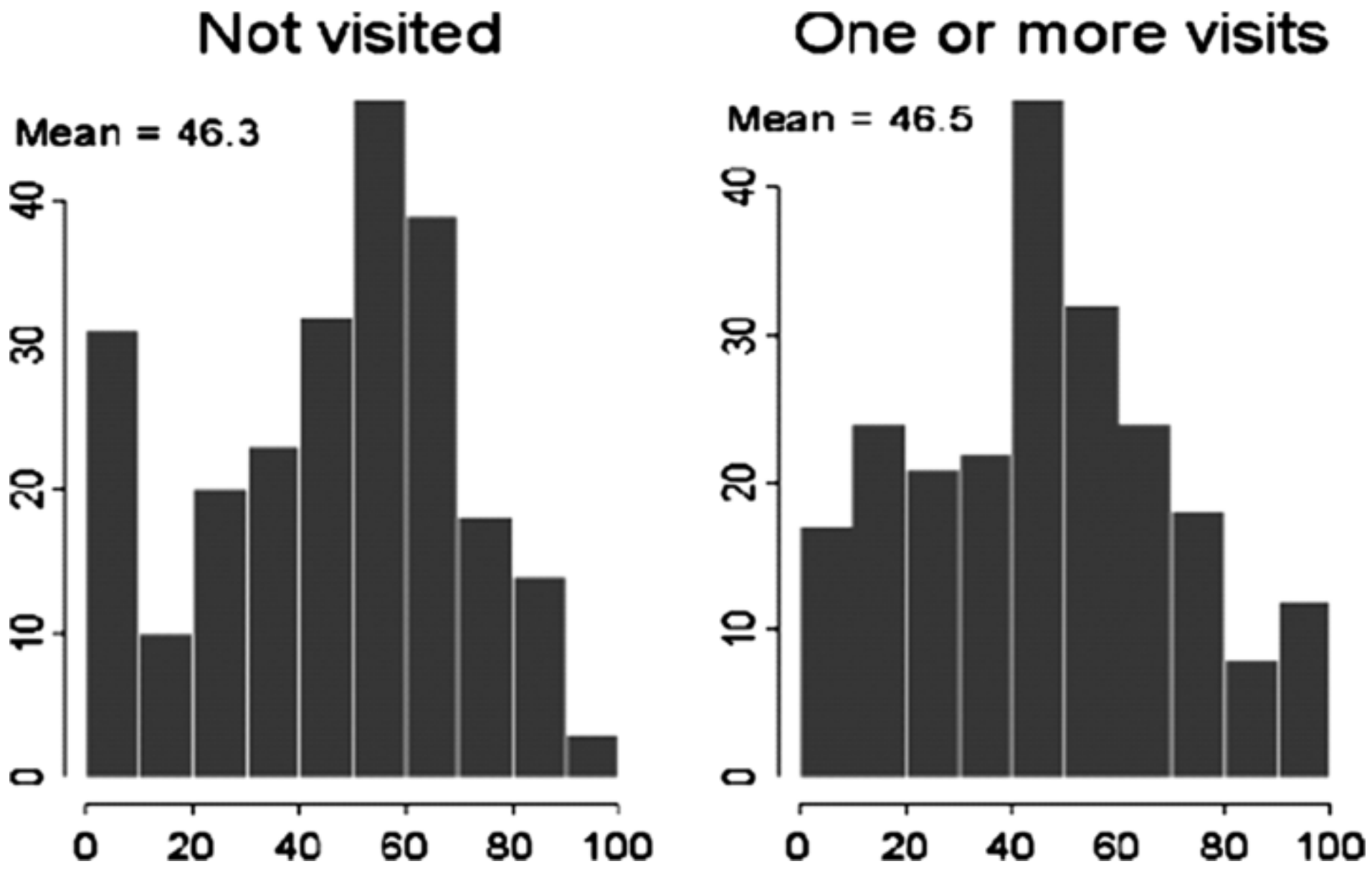

Fig. 4. *Histograms for a variable in a subclass of propensity scores: prior Rx score.*

In any case the design phase was complete, except for the specification of model-based adjustments to be made within the bins, and the more detailed analyses used to rank doctors by priority to visit. Readers interested in the conclusions, which are a bit surprising, should check Rubin and Waterman (2006).

## 5. A principal stratification example.

5.1. *The causal effect of being treated in large volume versus small volume hospitals.* The third example illustrates the point that the design phase in some observational studies may involve conceptualizing the hypothetical underlying randomized experiment that lead to the observed data as being more complex than a randomized block or randomized paired comparison. In particular, in some situations, we may have to view the hypothetical experiment as being a randomized block with noncompliance to the assigned treatment, a so-called “encouragement” design [Holland (1988)]. In many settings with human subjects, even an essentially perfectly designed randomized experiment only randomly assigns the encouragement to take treatment or control because we cannot force people to take one or the other. In the context of a perfectly double-blind experiment, where the subjects have no idea whether they are getting treatment or control, there will be no difference in compliance rates between the treatment versus control groups, but there are often side effects that create different levels of compliance in the conditions [Jin and Rubin (2008)]. In such cases, the ideas behind "instrumental variables" methods [Angrist, Imbens and Rubin (1996)] as generalized to "principal stratification" [Frangakis and Rubin (2002)] can be very useful.

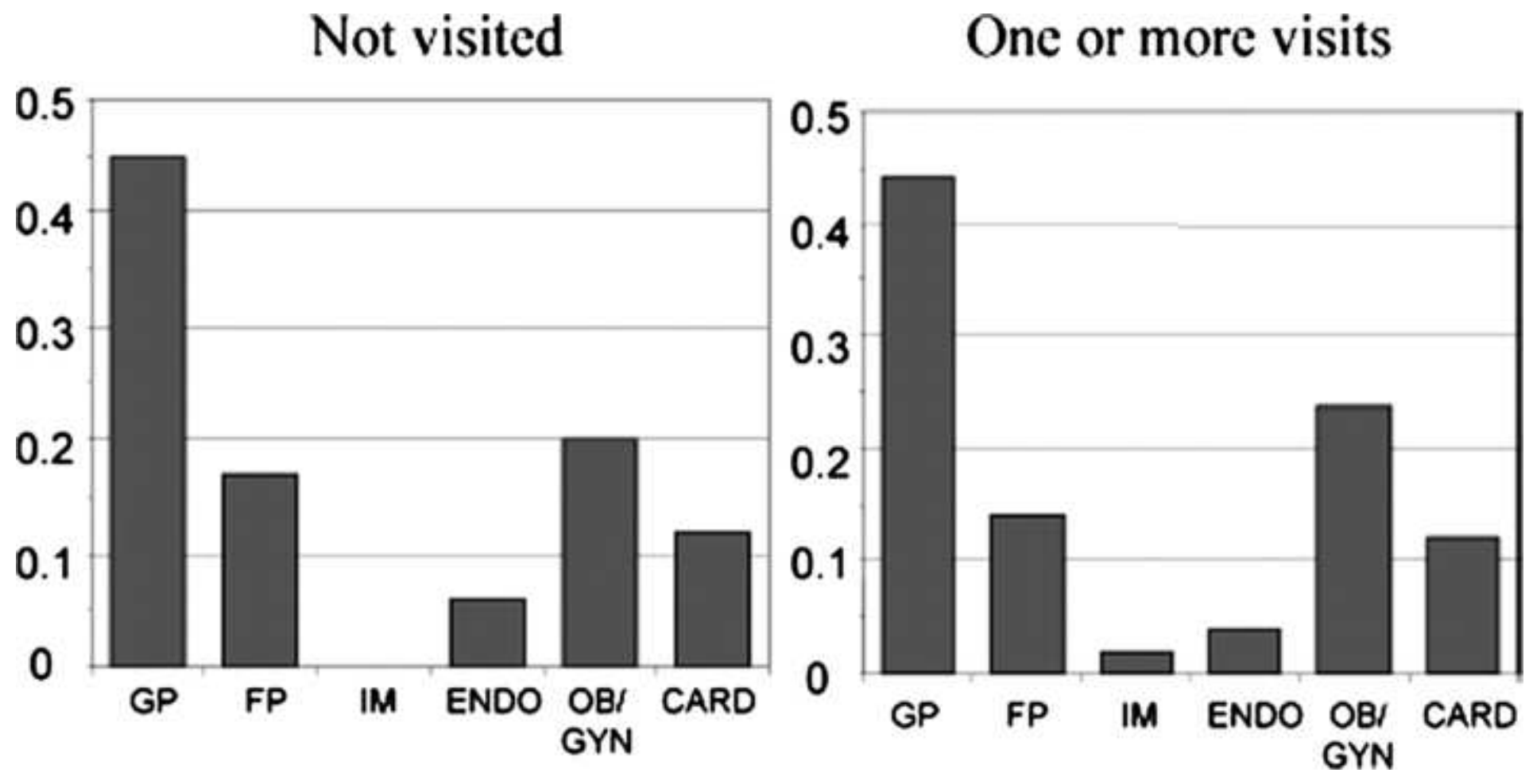

FIG. 5. *Histograms for a variable in a subclass of propensity scores: specialty.*

5.2. *Propensity score subclassification for diagnosing hospital type.* We illustrate this design using a small observational data set from the Karolinska Institute in Stockholm, Sweden. Interest focuses on the treatment of cardia cancer patients in Central and Northern Sweden, and whether it is better for these patients to be treated in a large or small volume hospital, where volume is defined by the number of patients with that type of cancer treated in recent years. The data set has 158 cardia cancer patients diagnosed between 1988 and 1995, 79 diagnosed at large volume hospitals, defined as treating more than ten patients with cardia cancer during that period, and 79 diagnosed at the remaining small volume hospitals. These sample sizes are small, but the data set is the only one currently available in Sweden to study this important question.

Generally, the commonly held view is that being treated in a large volume hospital is better, but the opposite argument could be made when the large volume treating hospital is far from a support system of family and friends, which presumably may be more available in small volume hospitals. The most critical policy issue concerns whether the cardia cancer treatment centers at small volume hospitals can be closed without having a deleterious effect on patient survival. If so, resources could be saved because patients diagnosed at small volume hospitals could be transferred to large volume treating hospitals, and if it is true that large volume cardia cancer treatment centers offer better survival outcomes, then the small volume ones should arguably be closed in any case. Our data set has hospital volume and patient survival information in it.

Because of the uniform training of doctors under the socialized medical system in Sweden, the assignment of large versus small "home hospital type," where the cancer was diagnosed, was considered by medical experts to be unconfounded, that is, essentially assigned at random within levels of measured covariates, $X$: age at diagnosis, date of diagnosis, sex of patient and urbanization. The decision maker is the individual patient, so our dataset seems well-suited for studying the causal effect of home hospital type on survival.

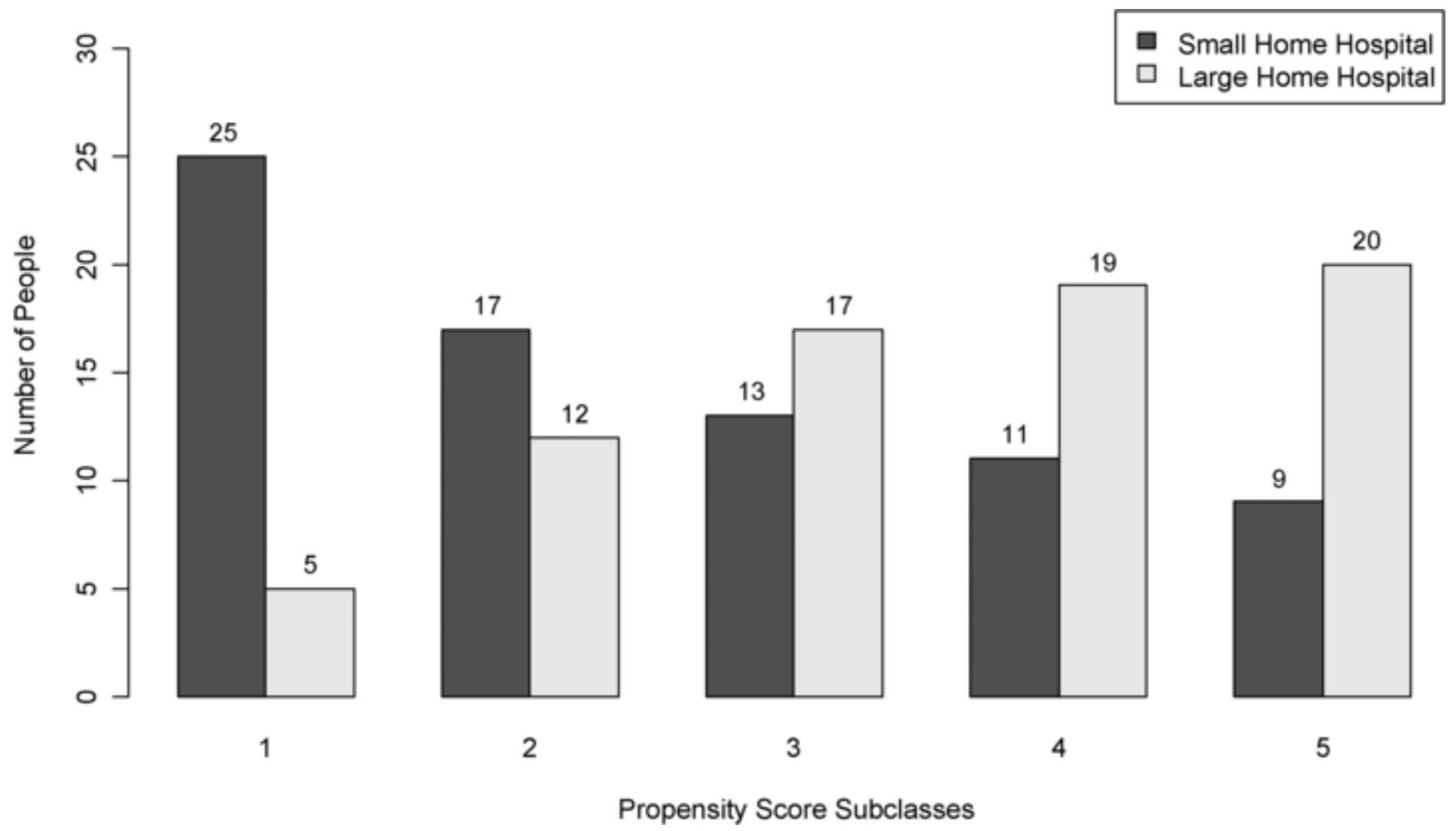

Fig. 6. *Cardia cancer, number of people, subclassified by propensity score.*

Propensity score analyses were done to predict diagnosing (home) hospital type from $X$, including nonlinear terms in $X$. It was decided that the age of patient should be limited to between 35 and 84 because the two patients under 35 (actually both under 30) were both diagnosed in large volume hospitals, and longer term survival in the 8 cardia cancer patients 85 and over was considered unlikely no matter where treated, and would therefore simply add noise to the survival data. Propensity score analyses on the remaining 148 patients led to five subclasses; these are summarized in Figures 6–8 are "Love plots" [Ahmed et al. (2006)] summarizing balance before and after this subclassification, for binary and continuous covariates, respectively.

5.3. *Treating hospital type versus home hospital type.* If patients were always treated in the same hospital where they were diagnosed, estimating the causal effects of hospital type would now be easy because of the assumed unconfounded assignment of diagnosing hospital type. However, there are transfers between hospital types, typically from small to large—33 of the 75 diagnoised in a small hospital transferred to a large one for treatment, but sometimes from large to small—2 of 75 transferred this direction. The reasons for these transfers are considered quite complex. The decisions are made by the individual patient, but clearly with input from doctors, relatives, and friends, where the issues being discussed include speculation about the probability of success of the treatment at one versus the other, the patient's willingness to tolerate invasive operations, the importance of being close to relatives and friends, and a host of other reasons. Consequently, there is no doubt that given the observed covariates and the home hospital type, the assignment of treating hospital type is confounded. Therefore, doing a direct analysis of treating hospital type, even if propensity score methods were used to create subclasses of patients with identical distributions of all observed covariates in large and small treating hospitals, would be considered unsatisfactory because key covariates were not available in the data set.

We can, however, still make progress based on the assumed unconfounded assignment of home hospital type by using a different template for our observational study of treating hospital type: a randomized experiment with noncompliance. That is, think of patients who transfer, or, more generally, who would have transferred if assigned to a different hospital type, as being noncompliers, and therefore, our template is that of a randomized encouragement design, where the encouragement to be treated in the diagnosing large or small hospital is randomly assigned within propensity score strata. The crucial idea here is then to stratify also on the bivariate "intermediate outcome," treating hospital when assigned to a large home hospital and treating hospital when assigned to a small home hospital. Even though only one of these intermediate variables is actually observed, progress can still be made. Notice that the design phase does here look at intermediate outcome data, treating hospital type, but not the outcome data on survival, on which decisions will be based. Survival data are not available at this stage!

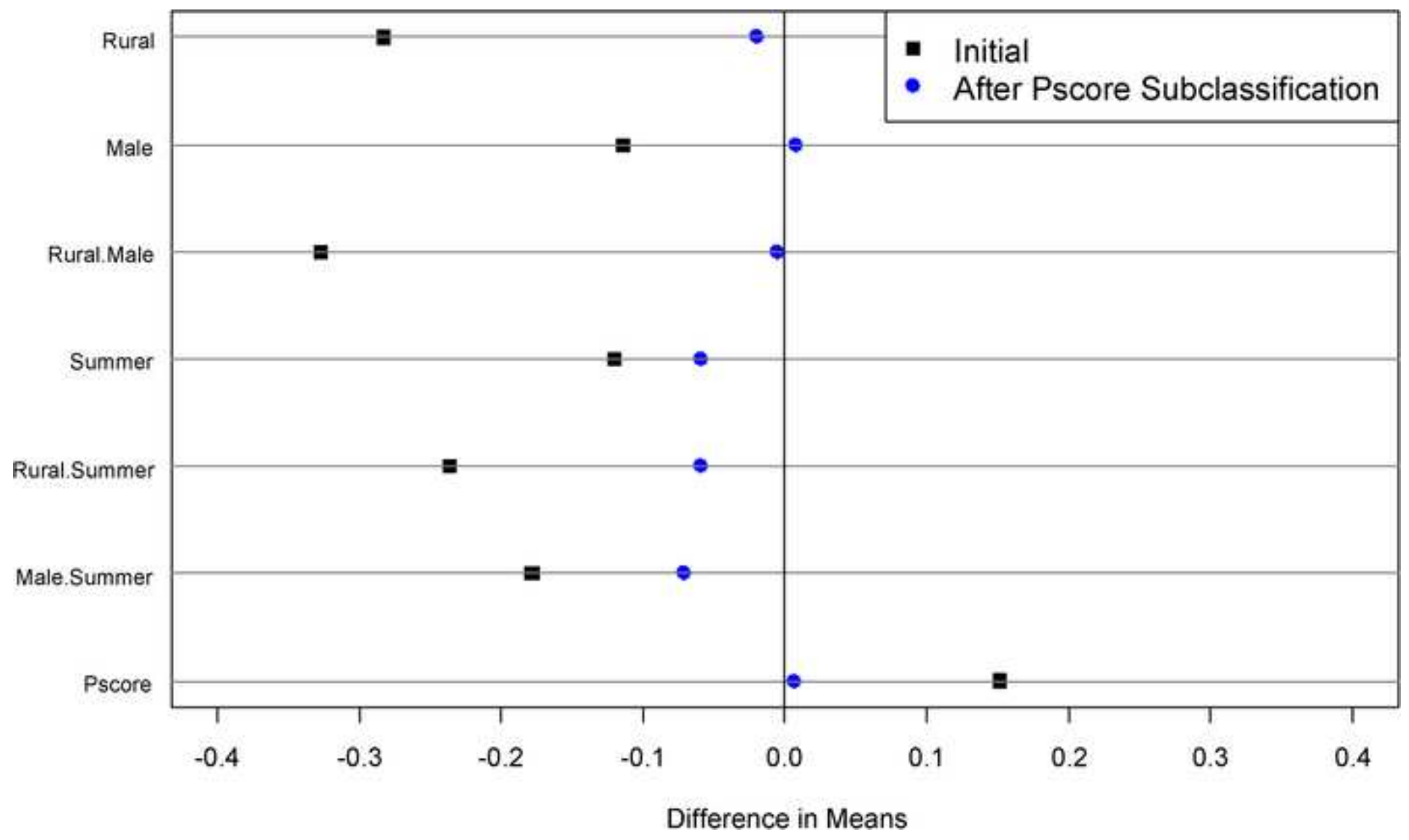

Fig. 7. *Cardia cancer, difference in means for binary covariates and pscore.*

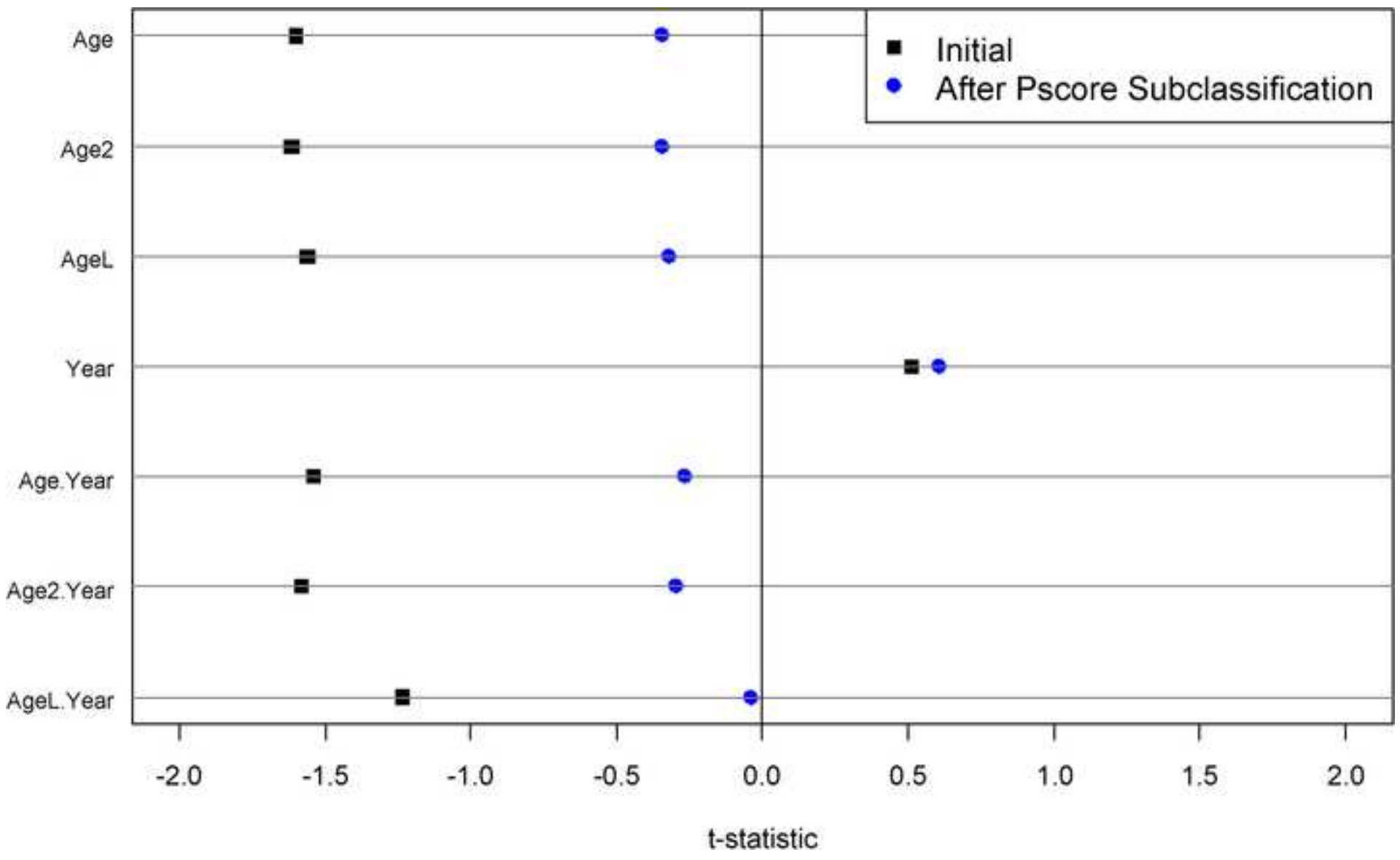

Fig. 8. *Cardia cancer, t-statistics for continuous covariates.*

Table 3
*Cardia cancer: observed counts in observed groups and approximate counts in principal strata under monotonicity assumption—subclass 1*

| | (1) | | (2) | (3) | (4) | | (5) | (6) |
|---|---|---|---|---|---|---|---|---|
| | **"Assigned"/ randomized home hospital type** | | **Treating hospital type $T$** | | **Underlying principal strata: $h =$** | | **Approximate proportion in population in principal strata** | **Approximate $N$ in LS principal stratum** |
| | $h$ | # | | # | $\ell$ | $s$ | | |
| (1) | $\ell$ | 5 | $L$ | 5 | $L$ | $L$ | 44% | |
| | | | | | $L$ | $S$ | 56% | 3 |
| (2) | | | $S$ | 0 | $S$ | $S$ | 0% | |
| (3) | $s$ | 25 | $L$ | 11 | $L$ | $L$ | 44% | |
| | | | | | $S$ | $S$ | 0% | |
| (4) | | | $S$ | 14 | $L$ | $S$ | 56% | 14 |

Denote the home hospital type by $h$, which takes the value $\ell$ when assigned large hospital type and $s$ when assigned small home hospital type. Similarly, let $T$ denote treating hospital type, which takes the value $L$ when the treating hospital is large, and takes the value $S$ when treating hospital is small. The first three columns of Tables 3–7 summarize the observed values of $h$ and $T$ within each of the five propensity score subclasses. Clearly, in all subclasses, transfers into large hospitals are common, but only in subclass 5 are there any $\ell \to S$ transfers. But do we estimate that there are compliers, who are treated in both large and small treating hospital types, within each subclass? If not, we will not be able to estimate the causal effect of treating hospital type for the entire group of patients—a critical design issue with this template.

TABLE 4

*Cardia cancer: observed counts in observed groups and approximate counts in principal strata under monotonicity assumption—subclass 2*

| | (1) | | (2) | (3) | (4) | | (5) | (6) |
|---|---|---|---|---|---|---|---|---|
| | **"Assigned"/ randomized home hospital type** | | **Treating hospital type $T$** | | **Underlying principal strata: $h =$** | | **Approximate proportion in population in principal strata** | **Approximate $N$ in LS principal stratum** |
| | $h$ | # | | # | $\ell$ | $s$ | | |
| (1) | $\ell$ | 12 | $L$ | 12 | $L$ | $L$ | 71% | 3 |
| | | | | | $L$ | $S$ | 29% | |
| (2) | | | $S$ | 0 | $S$ | $S$ | 0% | |
| (3) | $s$ | 17 | $L$ | 12 | $L$ | $L$ | 71% | |
| | | | | | $S$ | $S$ | 0% | |
| (4) | | | $S$ | 5 | $L$ | $S$ | 29% | 5 |

TABLE 5

*Cardia cancer: observed counts in observed groups and approximate counts in principal strata under monotonicity assumption—subclass 3*

| | (1) | | (2) | (3) | (4) | | (5) | (6) |
|---|---|---|---|---|---|---|---|---|
| | **"Assigned"/ randomized home hospital type** | | **Treating hospital type $T$** | | **Underlying principal strata: $h =$** | | **Approximate proportion in population in principal strata** | **Approximate $N$ in LS principal stratum** |
| | $h$ | # | | # | $\ell$ | $s$ | | |
| (1) | $\ell$ | 17 | $L$ | 17 | $L$ | $L$ | 38% | 11 |
| | | | | | $L$ | $S$ | 62% | |
| (2) | | | $S$ | 0 | $S$ | $S$ | 0% | |
| (3) | $s$ | 13 | $L$ | 5 | $L$ | $L$ | 38% | |
| | | | | | $S$ | $S$ | 0% | |
| (4) | | | $S$ | 8 | $L$ | $S$ | 62% | 8 |

5.4. *Principal strata and the monotonicity assumption.* Formally in the RCM, there are two types of outcomes: (1) the treating hospital type, $T$, which equals $T(\ell)$ when $h = \ell$ and $T(s)$ when $h = s$, and (2) the survival time since diagnosis, $Y$, which equals $Y(\ell)$ when $h = \ell$ and $Y(s)$ when $h = s$. The possible values of $(T(\ell), T(s))$ will be denoted $LL, LS, SL$, or $SS$ [where, for simplicity, $LL$ means the same as $(L, L)$, etc.], and those values define four possible "principal strata." $LS$ can be thought of as the stratum of compliers, that is, nontransfer patients; the $LL$ and $SS$ strata can be thought of as noncompliers who will always be treated at the same hospital type no matter where assigned, and $SL$ can be thought of as defiers, who will transfer no matter where assigned. The values of the principal strata are *not* affected by assignment of home hospital type—which value $[T(\ell)$ or $T(s)]$ is observed *is* affected by treatment assignment, but the bivariate values are not, and therefore $(T(\ell), T(s))$ is, formerly, a partially observed covariate.

TABLE 6
*Cardia cancer: observed counts in observed groups and approximate counts in principal strata under monotonicity assumption—subclass 4*

| | (1) | | (2) | (3) | (4) | | (5) | (6) |
|---|---|---|---|---|---|---|---|---|
| | **"Assigned"/ randomized home hospital type** | | **Treating hospital type $T$** | | **Underlying principal strata: $h =$** | | **Approximate proportion in population in principal strata** | **Approximate $N$ in LS principal stratum** |
| | $h$ | # | | # | $\ell$ | $s$ | | |
| (1) | $\ell$ | 19 | $L$ | 19 | $L$ | $L$ | 55% | |
| | | | | | $L$ | $S$ | 45% | 9 |
| (2) | | | $S$ | 0 | $S$ | $S$ | 0% | |
| (3) | $s$ | 11 | $L$ | 6 | $L$ | $L$ | 55% | |
| | | | | | $S$ | $S$ | 0% | |
| (4) | | | $S$ | 5 | $L$ | $S$ | 45% | 5 |

TABLE 7
*Cardia cancer: observed counts in observed groups and approximate counts in principal strata under monotonicity assumption—subclass 5*

| | (1) | | (2) | (3) | (4) | | (5) | (6) |
|---|---|---|---|---|---|---|---|---|
| | **"Assigned"/ randomized home hospital type** | | **Treating hospital type $T$** | | **Underlying principal strata: $h =$** | | **Approximate proportion in population in principal strata** | **Approximate $N$ in LS principal stratum** |
| | $h$ | # | | # | $\ell$ | $s$ | | |
| (1) | $\ell$ | 20 | $L$ | 18 | $L$ | $L$ | 67% | |
| | | | | | $L$ | $S$ | 23% | 5 |
| (2) | | | $S$ | 2 | $S$ | $S$ | 10% | |
| (3) | $s$ | 9 | $L$ | 6 | $L$ | $L$ | 67% | |
| | | | | | $S$ | $S$ | 10% | |
| (4) | | | $S$ | 3 | $L$ | $S$ | 23% | 2 |

Now, we consider what is called the "monotonicity" assumption or the "no-defier" assumption—that is, we assume that the $SL$ principal stratum is empty. In our setting, this assumption is very plausible, and because it excludes the $SL$ principal stratum, we have only three principal strata: $LL$, $LS$ and $SS$. Under this assumption, the possible principal strata for each *observed* combination of home hospital type and treating hospital type in each propensity subclass are shown in the fourth columns of Tables 3–7. The observed $\ell \to S$ group (the second row in Tables 3–7) must be composed of $SS$ patients because they can be neither $LL$ nor $SL$ patients, respectively—because they were assigned $\ell$ but treated in $S$ and therefore are not $LL$ patients, and there are no $SL$ patients by the monotonicity assumption. Similarly, the observed $s \to L$ group (the third row of Tables 3–7) must be $LL$ patients because they were assigned $s$ but were treated in $L$.

In contrast, the observed $\ell \to L$ subgroup (the first row of Tables 3–7) could be compliers, and so be in $LS$, or noncompliers who are members of the $LL$ principal stratum (who were assigned to home hospital type $L$, and to which they would have transferred for their treating hospital type if they were assigned to a small home hospital type). Hence, we split row 1 into two sub-rows in the fourth column of Tables 3–7. Similarly, the observed $s \to S$ subgroups (the fourth row of Tables 3–7) could be compliers, and so be in $LS$, or noncompliers who are members of the $SS$ principal stratum, and so is also split into two sub-rows.

We can approximate the proportion of patients in each principal stratum, as shown in the fifth columns of Tables 3–7. More explicitly, from the second row of Table 7, columns (1) and (3), we see that 2/20 are observed to be $\ell \to S$. Because of the assumed random assignment into $\ell$ and $s$ within propensity score subclasses, we have that approximately 10% of the patients belong to the principal stratum $SS$, as shown in the fifth column of Table 7. Similarly, from the third row of Table 7, columns (1) and (3), we infer that approximately $6/9 \approx 67\%$ of patients belong to principal stratum $LL$ in this subclass, as shown in the fifth columns of Table 7.

Hence, we can approximate the fraction of compliers, the $LS$ principal stratum in this subclass, by simple subtraction: $100\% - 10\% - 67\% = 23\%$. The sixth column in Table 7 indicates the approximate number of $LS$ patients in each of the four rows of observed patients. Analogous calculations are summarized in Tables 3–6 for the other propensity score subclasses. Even if we could perfectly identify all the $LS$ patients, which we cannot, the sample sizes are small, and so inference for the causal effect of treating hospital when it equals home hospital will be imprecise. Nevertheless, we outline the planned analysis in Section 5.6 because these are the only data available to study this question. Importantly, we anticipate that in each subclass there are some compliers who are treated in large volume hospitals and some compliers who are treated in small volume hospitals.

5.5. *ITT and CACE* $= ITT_{LS}$ *and their estimation.* The average causal effect of home hospital type on survival is the comparison of the potential survival outcomes of all $N$ patients under $h_i = \ell$ and under $h_i = s$,

$$\mathrm{ITT} = \frac{1}{N}\sum_{i=1}^{N}[Y_i(\ell) - Y_i(s)],$$

where ITT is the Intention-To-Treat (ITT) effect of the assignment of large versus small home hospital type. Under unconfounded assignment of home hospital type, we are able to estimate ITT by taking the average observed difference in $Y$ for large volume hospital patients and small volume hospital patients within each propensity subclass, weighting each subclass-specific estimate by the total number in that subclass and averaging the estimates. Because we are not examining survival outcome data at this design stage, we cannot calculate these estimates, but we saw this approach in the breast cancer example of Section 4. In this problem using the template of a randomized block experiment with noncompiliance, the estimation is more subtle.

When $G_i = LS$, the home hospital type equals the treating hospital type, that is, $h_i = T_i$. The causal effect of home hospital type in the $LS$ principal stratum is defined to be

$$\mathrm{CACE} \equiv \mathrm{ITT}_{LS} = \frac{1}{N_{LS}}\sum_{i\in LS}(Y_i(L) - Y_i(S)),$$

where $N_{LS}$ is the number of $LS$ patients, and CACE means "Compliance Average Causal Effect" [Imbens and Rubin (1997)]. $\mathrm{ITT}_{LS}$ can be interpreted as either the intention-to-treat effect of home hospital type for complying patients or the intention-to-treat effect of treating hospital type for complying patients, because for the $LS$ principal stratum, $h_i = T_i$. Under monotonicity, the $LS$ principal stratum is the only stratum of patients where we can learn about the causal effects of treating hospital type because the patients in the other principal strata, $LL$ and $SS$, will always be exposed to the same treating hospital type.

CACE is easily estimated once we identify the individuals in the $LS$ stratum, and we have not yet identified any particular member of the $\ell \to L$ or $s \to S$ rows (rows one and four) in Tables 3–7 as being in the $LS$ principal stratum, and so we cannot yet compare average outcomes in this stratum. Nevertheless, we can find a unique method-of-moments estimate of the causal effects of assigned (= treating) hospital type within the $LS$ principal stratum under, what are considered, medically very justifiable assumptions, which in general are called "exclusion restrictions." The resulting estimator of CACE is known as the "instrumental variable estimate" [Angrist, Imbens and Rubin (1996)]. Better (e.g., Bayesian) methods of estimation exist [e.g., see Imbens and Rubin (1997)].

5.6. *Exclusion restrictions.* There are two exclusions restrictions. The first exclusion restriction is for patients in the $LL$ principal stratum. It states that, for all $i \in LL, Y_i(\ell) = Y_i(s)$, that is, there is no effect on potential outcomes $Y$ of being assigned to a large ($\ell$) versus small ($s$) home hospital type for patient $i \in LL$. The medical justification for this restriction is that patient $i$ would be treated in a large hospital type ($L$) under either assignment, and one's medical outcome is considered a result of where one is treated not where one is diagnosed. The exclusion restriction for patients in the $SS$ principal stratum is analogous; for all $i \in SS, Y_i(\ell) = Y_i(s)$, that is, for those patients who would be treated in a small hospital type ($S$) whether assigned to $l$ or $s$, there is no effect of assignment on the $Y$ potential outcomes.

Now, ITT for all patients can be written as

$$\mathrm{ITT} = \pi_{LS}\mathrm{ITT}_{LS} + \pi_{SS}\mathrm{ITT}_{SS} + \pi_{LL}\mathrm{ITT}_{LL},$$

where $\pi_{LS}, \pi_{SS}$ and $\pi_{LL}$ are the fractions of the sample, and $\mathrm{ITT}_{LS}$, $\mathrm{ITT}_{SS}$ and $\mathrm{ITT}_{LL}$ are the intention-to-treat effects in the $LS$, $SS$ and $LL$ strata, respectively. Because the exclusion restrictions force $\mathrm{ITT}_{SS}$ and $\mathrm{ITT}_{LL}$ to be identically zero, this equation becomes

$$\mathrm{ITT} = \pi_{LS}\mathrm{ITT}_{LS},$$

or

$$\mathrm{ITT}_{LS} = \mathrm{ITT}/\pi_{LS}.$$

Thus, the instrumental variables estimate of the ITT effect of treating hospital among compliers is found by dividing the estimated ITT effect of home hospital type by the estimated fraction of the sample in $LS$.

The planned analysis will use Bayesian versions of this estimator within each propensity score subclass, and then average over all subclasses. An initial Bayesian analysis that partially benefits from the propensity score analysis presented here, but also involves data on stomach cancer patients, is presented in Rubin et al. (2008).

**6. Discussion.** This article advocates the position that observational studies for causal effects need to be designed to approximate randomized experiments. This enterprise requires careful thought and execution, and not simply running mindless regression programs and looking at coefficients. In most situations, this design effort will be more intellectually demanding than a similar effort for an analogous randomized experiment. Of critical importance, final outcome data cannot be used in design without compromising the objectivity of the study design. Propensity score methods are extremely helpful tools for reconstructing the underlying hypothetical experiment that lead to the observed data. Sometimes, the hypothetical approximating randomized experiment is one with complications, such as noncompliance, and then the principal stratification framework can be extremely helpful. But most important is for the worker in observational studies to stay focused on approximating a plausible hypothetical underlying randomized experiment.

A final comment concerns the application of this perspective to actual randomized experiments, especially those with covariates that have not been used in the randomization (e.g., not used to create blocks). In such cases, we would expect random imbalances in some covariates, and if there is concern that these covariates may be related to outcomes, the application of the techniques described here, with no access to final outcome data, preserves the objectivity of the experiment, whereas model-based adjustments, unless fully specified a priori, would compromise that objectivity. This approach has been applied, for example, in a study of school vouchers [Barnard et al. (2003)] and in a study of vertical disease transmission during delivery [Zell et al. (2007)].

**Acknowledgments.** Thanks to Cassandra Wolos for the propensity score analyses of Section 5.2, to Elizabeth Zell for very constructive comments on various drafts, to the editorial board for very helpful comments and a wonderfully rapid review, and finally to two outside reviewers for generous and helpful comments.

DEPARTMENT OF STATISTICS
HARVARD UNIVERSITY
CAMBRIDGE, MASSACHUSETTS 02138
USA
E-MAIL: rubin@stat.harvard.edu